\documentclass{iaus}
\usepackage{graphicx}

\title[The post-nova population] %% give here short title %%
{The post-nova population}

\author[Linda Schmidtobreick et al.]   %% give here short author list %%
{Linda~Schmidtobreick$^1$,
Claus~Tappert$^2$,
Alessandro~Ederoclite$^3$,
Nikolaus~Vogt$^2$}

\affiliation{$^1$European Southern Observatory, Alonso de Cordova 3107, Vitacura, Santiago, Chile\\email: {\tt lschmidt@eso.org}\\[\affilskip]
$^2$Dept. de F\'\i sica y Astronom\'\i a, Universidad de Valpara\'\i so, Av. Gran Breta\~n a 1111, Valpara\'\i so, Chile\\[\affilskip]
$^3$ Instituto de Astrof\'\i sica de Canarias, 38200, La Laguna, Tenerife, Spain}

\pubyear{2011}
\volume{xxx}  %% insert here IAU Symposium No.
\pagerange{xxx}
\jname{xxx}
\editors{A.C. Editor, B.D. Editor \& C.E. Editor, eds.}
\begin{document}

\maketitle

\begin{abstract}
We here present  our on-going project to unveil
the post-nova population by re-discovering old novae that have 
been lost after the initial outburst and of which the binary itself 
is unobserved. We take UBVR photometry for the candidate
selection, long-slit spectroscopy to confirm these candidates, and
time-resolved spectroscopy to measure the orbital period of the newly
confirmed post-novae. Some preliminary results are shown as examples.
\end{abstract}

\firstsection % if your document starts with a section,
\section{Introduction}
Classical novae are a subgroup of cataclysmic variables that went through
a thermonuclear runaway explosion on the white dwarf. These white dwarfs
are considered possible candidates for Type Ia supernovae. Especially
recurrent novae are known to exhibit massive white dwarfs close to the
Chandrasekhar limit and at the same time have high accretion rates, both
ingredients to make a future SN explosion likely. Classical
novae with only one observed nova outburst are more frequent but their
binary parameters like e.g. the white dwarf mass are less well studied.
To do this, one has to wait a few decades after the nova eruption
until the characteristics of the underlying CV become dominant in the
post-nova (see e.g. the detailed studies of RR\,Pic by \cite[Schmidtobreick et al. 2003a]{schmidto+03a} and \cite[2008]{schmidto+08}). 
Generally, the long-term behaviour of pre- and postnovae is an open 
question (\cite[Vogt 1990]{vogt90}; \cite[Duerbeck 1992]{duerbeck92}) but important -- for instance 
in context with the hibernation scenario (\cite[Shara et al. 1986]{shara+86}). 
On the other hand, our actual knowledge of post-novae is largely
incomplete: almost 3/4 of the 199 nova candidates that erupted before
1980 still lack spectroscopic observations or even an unambiguous
identification of the post-nova.

A few years ago, we have started a program to re-discover old novae
but concentrating on the ones which exhibited large outburst amplitude.
We recovered and confirmed several of these systems (see e.g.\
\cite[Schmidtobreick et al. 2003b]{schmidtobreick+03b},
\cite[2005]{schmidtobreick+05}). 
We now use a similar approach to find general old novae --
not necessarily high outburst amplitude ones. And we go one step further
in the sense that we do not only want to recover the systems, but we
also intend to derive the basic properties of the binaries 
%and for this reason need to
which includes at the very least the determination of the
%at least determine the
orbital period of the recovered systems.

\section{The method}
\begin{figure}[t]
\rotatebox{-90}{\includegraphics[width=4.7cm]{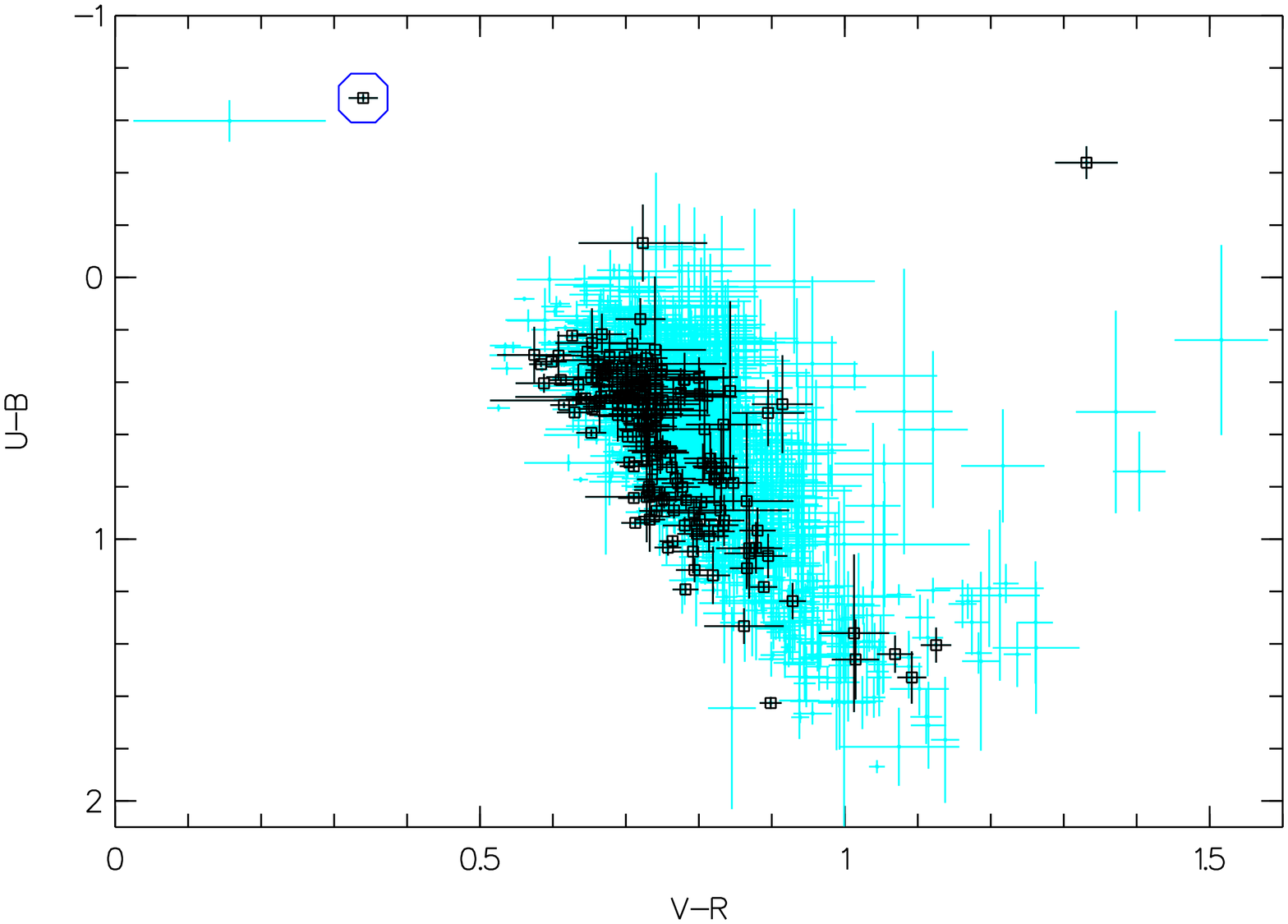}}~
\rotatebox{-90}{\includegraphics[width=4.7cm]{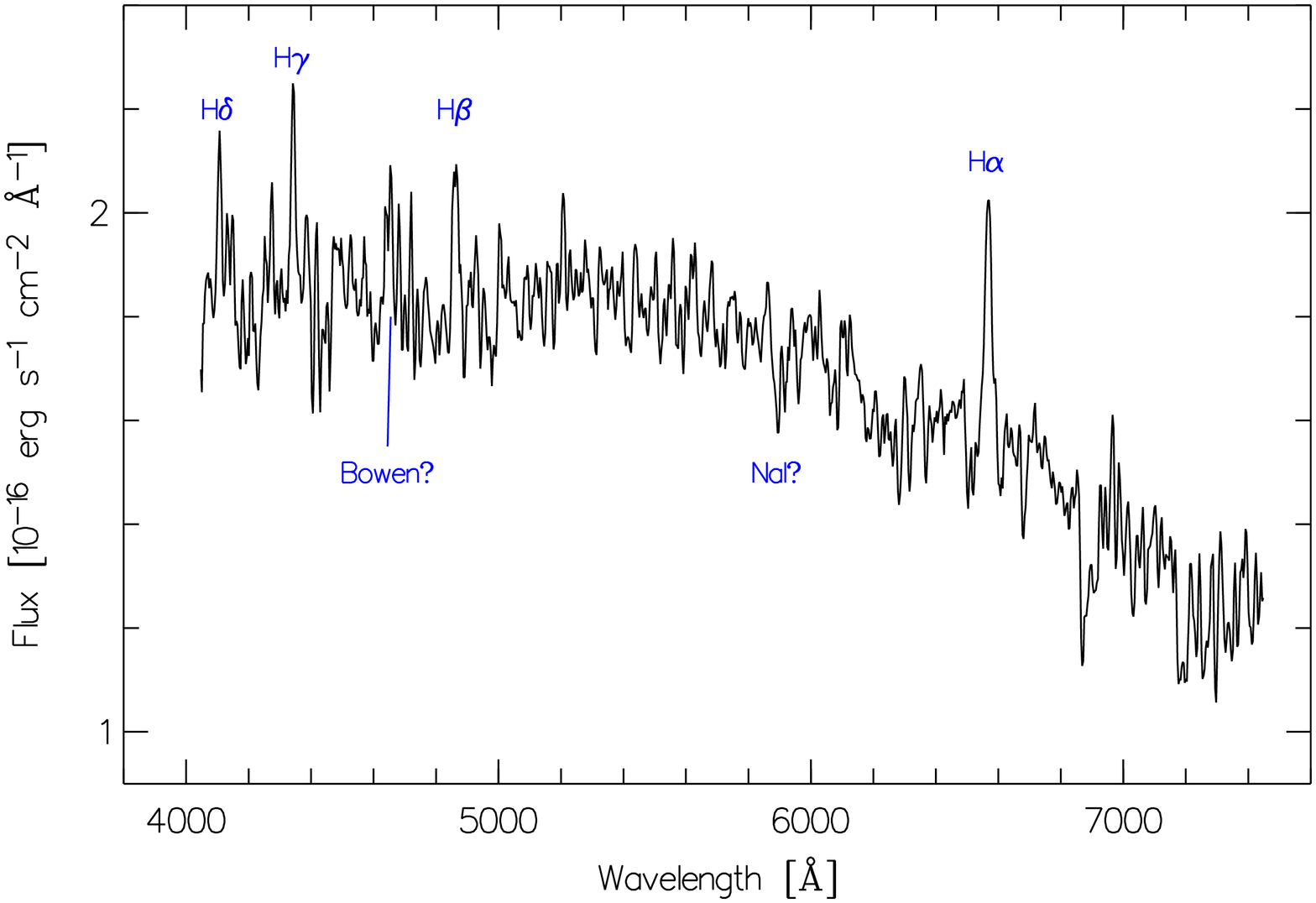}}
 \caption{\label{ilnor} The left plot shows the colour-colour diagram
of all stars in the field surrounding IL\,Nor -- the stars in the inner $300\times 300$ pixels are plotted as black squares. The best candidate for the
old nova is marked with a circle. The right plot shows the spectrum of this
best candidate. The typical emission lines confirm the classification as a cataclysmic variable.}
\end{figure}

The observations are  done in three steps:\\
(1) To select possible candidates for the old novae, we make use of the fact
that the light from a cataclysmic variable is provided by at least three 
different physical components: the white dwarf, the secondary star, and the
accretion disc or stream. Due to its size and high temperature, the latter 
is generally the dominant source in the optical range. 
While the white dwarf contribution
affects mainly the UV range of the spectrum, the secondary late--type star
contributes on the red and infrared side.
The compound of these light sources results in characteristic colour terms
that are distinct from normal field stars. Cataclysmic variables appear
as generally very blue objects with a shift towards the red
at longer wavelengths. In a colour--colour diagram,
they are thus found on the blue side but slightly above the main sequence.
As an example, see the left diagram of Figure\,\ref{ilnor}.\\
(2) To confirm the candidates of step (1), we take low resolution
spectra and identify the cataclysmic variable via the
typical emission lines (see right diagram of Figure\,\ref{ilnor}).\\
(3) The last step is to take medium resolution time series spectroscopy
of the confirmed old nova. We use the radial velocity variation of strong emission lines like H$\alpha$ to determine at least the orbital period. In case
of good S/N and coverage, also a further study of the line profile variation and a
corresponding analysis of the accretion process
is possible.
\section{The sample so far}
In total, we have observations for 21 novae so far. 17 systems have 
been confirmed spectroscopically, for the remaining four, several 
possible candidates were identified and await the spectroscopic
confirmation. We have determined the orbital period for 
three systems and have constraints for two more. For more details on 
the results and a thorough analysis, we refer to 
\cite[Tappert et al. (2011)]{tappert+11}. We conclude that our
method works very well and the efficiency to recover the old novae is
high. 
%To get a sufficient number of systems for population studies,
%we are awaiting further observations.
Further observations are planned to obtain a sufficiently high number of 
systems for population studies.

\end{document}